\documentclass[12pt,a4paper]{article}
\usepackage{graphicx,psfrag,amsmath,amssymb,amsfonts,latexsym,color,bm,bbold}
\usepackage[colorlinks=true, linkcolor=blue, citecolor=blue, 
urlcolor=blue]{hyperref}
\usepackage[english]{babel}
\usepackage[utf8]{inputenc}
\usepackage{slashed}
\usepackage{physics}
\usepackage{tikz}
\usepackage{pgfplots}
\pgfplotsset{compat=1.18}
\usepackage[margin=2.5cm]{geometry}
\usepackage{pgfplotstable}
\begin{document}
\title{Parity-dependent Casimir forces and Hall currents for a confined Dirac 
field} 
\author{Aitor Fern\'andez and C\'esar D.~Fosco\\
{\normalsize\it Centro At\'omico Bariloche and Instituto Balseiro}\\
{\normalsize\it CNEA and UNCuyo}\\
{\normalsize\it R8402AGP S.\ C.\ de Bariloche, Argentina.} }
\maketitle
%====================================================================
\begin{abstract}
We study a massless Dirac field subjected to two alternative boundary 
conditions on two parallel thin walls, in $d+1$ dimensions. The 
two configurations correspond to the system being even or odd under reflection 
about the midplane between the two walls, and lead to qualitatively 
different behaviors.
The even (symmetric) configuration produces an attractive Casimir force, 
whereas the odd (antisymmetric) one yields repulsion, in agreement with a 
general theorem linking parity to the sign of the fermionic Casimir effect.

We complement this result by studying two phenomena associated with the vacuum fluctuations responsible for the Casimir interaction, both of which are also sensitive to parity: the correlation between 
currents concentrated on the walls, and the induced bulk current under 
the influence of an external electric field. For the latter we show that, in 
$2+1$ dimensions, an induced transverse (Hall-like) current arises, whose 
spatial profile inherits the symmetry of the confining potential.
\end{abstract}
%==========================================================================
%%%%%%%%%%%%%%%%%%%%%%%%%%%%%%%%%%%%%%%%%%%%%%%%%%%%%%%%%%%%%%%%%%%%%%%%%%%
%%%%%%%%%%%%%%%%%%%%%%%%%%%%%%%%%%%%%%%%%%%%%%%%%%%%%%%%%%%%%%%%%%%%%%%%%%%
%%%%%%%%%%%%%%%%%%%%%%%%%%%%%% Introduction %%%%%%%%%%%%%%%%%%%%%%%%%%%%%%% 
%%%%%%%%%%%%%%%%%%%%%%%%%%%%%%%%%%%%%%%%%%%%%%%%%%%%%%%%%%%%%%%%%%%%%%%%%%%
%%%%%%%%%%%%%%%%%%%%%%%%%%%%%%%%%%%%%%%%%%%%%%%%%%%%%%%%%%%%%%%%%%%%%%%%%%%
\section{Introduction}\label{sec:intro}
In the presence of non-trivial boundary conditions, the vacuum fluctuations 
of a quantum field may give rise to macroscopic manifestations, with the 
Casimir~\cite{Bordag2001,books} and related effects being among the most 
distinctive ones.

Except for very special cases, like highly diluted 
media~\cite{Dzyaloshinskii1961,Pitaevskii2008,Golestanian2009}, the Casimir 
force between two bodies cannot be obtained as the result of a 
superposition~\cite{RoLo2009,Milton2015,Paladugu2016}. This,
in turn, implies that the relationship between that force and the geometry 
(shapes and positions) of the intervening bodies, becomes generally rather 
involved.

In this context, what is perhaps the most relevant property of the effect is 
its {\em sign}, namely, whether it is an attractive or repulsive force. An 
important theoretical result~\cite{Kenneth:2006vr} relates this property to the 
behavior under a  parity transformation of the system. Indeed, it states that a 
sufficient condition for the force between two dielectric or conducting objects 
to be attractive is the existence of a reflection symmetry with respect to a 
plane. 
In a recent work~\cite{Fernandez:2025fpf}, we have studied a Dirac field in the 
presence of a background having support on two disconnected spatial regions,
interpreted as lumps of matter that, in appropriate limits, impose bag model boundary 
conditions~\cite{Chodos1974,Johnson1975,Arrizabalaga2017} on their respective 
frontiers.
The main result was that if a background is odd under reflection about the midplane between the two objects, they repel. 

In this work, we provide a concrete study of the consequences, for some 
relevant physical observables, of the system being even or odd under such a 
reflection. The model consists of two thin walls, each one of them imposing 
boundary conditions on its two faces. Since this leaves open the possibility of symmetric and antisymmetric setups, it is natural to compare their physical properties. 
In particular, we study the correlator between current operators concentrated 
on the walls, and we also study the vacuum currents, inside the slab, induced 
by an external, constant, uniform electric field which is perpendicular to the 
planes of the walls.

This paper is organized as follows: in Sect.~\ref{sec:system}, we define the 
physical system, including the action, the singular potential, and the 
conventions adopted. In Sect.~\ref{sec:Casimir}, we review the form of the 
Casimir energy for the two parity configurations and also show how the Casimir 
pressure for this system may be expressed in terms of the vacuum expectation 
value of a surface current operator, i.e., concentrated on one of the two 
walls. This is indeed even more clear in the case of $2+1$ dimensions.
 Based on this feature, in Sect.~\ref{sec:correlation} we deal with 
the vacuum correlation function between those surface currents at the two 
walls, an object which measures the effect of simultaneous deformations of the 
two boundaries.
In Sect.~\ref{sec:induced}, we compute the current induced by an external 
electric field perpendicular to the walls, and discuss its relation to the 
parity anomaly in $2+1$ dimensions. Results for the full fermionic propagator 
in the presence of one and two walls, required for our concrete calculations, 
are presented in Appendix~\ref{app:propagator}. The generalization of results
for the Casimir energy to the case of singular potentials with more general 
coupling constants is commented in Appendix B.
 
Finally, in Sect.~\ref{sec:Conclusions} we present a general discussion of the 
results obtained, and our conclusions.

%====================================================================
%%%%%%%%%%%%%%%%%%%%%%%%%%%%%%%%%%%%%%%%%%%%%%%%%%%%%%%%%%%%%%%%%%%%%
%%%%%%%%%%%%%%%%%%%%%%%%%%%%%%%%%%%%%%%%%%%%%%%%%%%%%%%%%%%%%%%%%%%%%
%%%%%%%%%%%%%%%%%%%%%%%%%%% The system %%%%%%%%%%%%%%%%%%%%%%%%%%%%%%
%%%%%%%%%%%%%%%%%%%%%%%%%%%%%%%%%%%%%%%%%%%%%%%%%%%%%%%%%%%%%%%%%%%%%
%%%%%%%%%%%%%%%%%%%%%%%%%%%%%%%%%%%%%%%%%%%%%%%%%%%%%%%%%%%%%%%%%%%%%
\section{The model}\label{sec:system}
We begin by defining the real-time action we shall consider in this paper;
it describes a Dirac field $\psi, \,\bar\psi$, including its coupling to an 
external gauge field $A$, and a ``potential'' term $V$, used to impose the
boundary conditions:
\begin{equation}\label{eq:action}
\mathcal{S}(\psi,\bar\psi; V, A) \;=\; \int d^nx~\bar\psi(x)
 \big[ i \not \!\! D \,-\, V(x^d) \big] \psi(x),
\end{equation}
where $n=d+1$ is the number of space-time dimensions, with 
$x=(x_\shortparallel,x^d)$ denoting the coordinates. Here, we 
have used $x_\shortparallel$ as a shorthand notation for $x^0, x^1, \ldots, 
x^{d-1}$, natural space-time coordinates for the constant-$x^d$ planes, where 
the boundary conditions are imposed. Indices from the beginning of the Greek 
alphabet shall be used for indices labeling the first $d$ space-time indices.

To that end we shall assume  the form of 
$V$ to be as follows:
\begin{equation}\label{eq:defv}
V(x^d) \;=\; g_L \,\delta(x^d-a_L) \,+\, g_R \,\delta(x^d-a_R) \;.
\end{equation}
The dimensionless constants $g_{L,R}$ characterize the strength of the 
singular interactions and $a_{L,R}$ the positions of the parallel 
infinite  walls, all perpendicular to the $x^d$ axis. This singular potential 
is used for imposing MIT bag boundary conditions on each side of each 
(zero-width) plane, as we explain at the end.

The covariant derivative: $D_\mu \equiv \partial_\mu + i e A_\mu$ introduces 
the minimal coupling with an electric charge $e$ to an electromagnetic field, 
while, in our conventions, the $\gamma$-matrices satisfy 
\mbox{$(\gamma^\mu)^\dagger=\gamma^0\gamma^\mu\gamma^0$} 
and $\left\{\gamma^\mu,\gamma^\nu\right\}=2 g^{\mu\nu}$, with 
$g^{\mu\nu}=\text{diag}(+1,-1,\dots,-1)$ the inverse metric (we have followed 
the general usage of not writing explicitly the
$2^{\lfloor\frac{n}{2}\rfloor}\times2^{\lfloor\frac{n}{2}\rfloor}$ identity 
matrix on the right hand side). The fermions are, for all $n$, in an 
irreducible representation of Clifford algebra. 

As a matter of convenience, we shall use a shorter notation for cases where
the external gauge field is absent; namely, 
$\mathcal{S}(\psi,\bar\psi; V)$ for $\mathcal{S}(\psi,\bar\psi; V, 0)$.

Let us now discuss the boundary conditions: as already mentioned, the system we 
consider contains fermions, and there is a slab, defined by two parallel 
planes which impose MIT bag boundary conditions. We 
recall that, for a static surface $\Sigma$, those conditions are usually 
implemented as follows:
\begin{equation}\label{eq:MITBBC}
	\Big(1+i\,\hat{n}(x) \cdot \vec{\gamma}\Big)\psi(x)\eval_{x\in\Sigma} 
	\,=\, 0 \;,
\end{equation}
where $\hat{n}(x)$ is a unit normal vector to $\Sigma$. It is worth noting 
that, in the absence of singularities in the field and in the equations it 
satisfies, both the inner and outer normal vectors do the job of leading to the 
vanishing of the normal current. Namely, 
\begin{equation}	
	n_\mu(x) \,\bar\psi(x)\gamma^\mu\psi(x)\eval_{x\in\Sigma}\;=\; 
	0,~~\text{with} \;\; n^\mu(x) \;\equiv\; \big(0,\hat{n}(x)\big) \;.
\end{equation}
In the case of the bag model, there is usually a natural choice, corresponding 
to the normal pointing towards the exterior of the bag. The reason for that is 
the following: assume a bag for a spatial region $U$ is implemented by a 
``potential well'' $V_U({\mathbf x})$ which in the conventions of our model 
corresponds to 
\begin{equation}
	V_U({\mathbf x}) \;\equiv\; V_0 \, \theta_U({\mathbf x}) 
\end{equation}
where $\theta_U({\mathbf x})$ vanishes when ${\mathbf x} \in U$ and equals $1$
otherwise. It is then rather straightforward to see that, when $V_0 \to + 
\infty$ the only remaining solutions must satisfy the bag condition with a 
normal which points outwards of $U$. Conversely, the inner normal 
appears when  $V_0 \to -\infty$. Both lead to a vanishing current on the 
boundary, and are both compatible with the self-adjointness of the Dirac 
Hamiltonian in the bag, as explained in~\cite{Arrizabalaga2017}. 

Here, however, we consider surfaces which impose boundary conditions on both 
of their faces. These may be generated by means of a singular potential 
localized on the surface. Indeed, for the $x^d=0$ plane, we consider the Dirac 
equation with a potential
\begin{equation}
	V(x^d) \,=\, g \delta(x^d)\,.
\end{equation}
By an immediate generalization to $d$ dimensions of the analysis presented 
in~\cite{Fosco2022,Fosco2023}, we get:
\begin{equation}
i\gamma^d\big[\psi(x_\shortparallel,\epsilon)-\psi(x_\shortparallel,-\epsilon) 
\big]-\frac{g}{2}\Big[\psi(x_\shortparallel,\epsilon)+\psi(x_\shortparallel,
-\epsilon)\Big]=0 \;.
\end{equation}

Defining the projectors
\begin{equation}
	\mathcal{P}_\pm\equiv\frac{1}{2}\left(\mathbb{1}\pm i\gamma^d\right) \;,
\end{equation}
one gets, in the special cases $g = \pm 2$, the conditions:
\begin{equation}
	\begin{cases}
		\mathcal{P}_\pm\psi(x_\shortparallel,\mp\epsilon)=0&\text{if }g=2\\
		\mathcal{P}_\pm\psi(x_\shortparallel,\pm\epsilon)=0&\text{if }g=-2,
	\end{cases}
\end{equation}
which correspond to bag boundary conditions with normal vector 
$\hat{n}(x)=(\vec{0}_\shortparallel,\pm1)$ at $x^d=\pm\epsilon$ (with the signs 
consistently chosen for each side of the wall). This agrees with the 
explanation in~\cite{Arrizabalaga2017} since, at least for a potential which 
approximates the $\delta$, when considering its two ``boundaries'' at 
$\pm\epsilon$, one obtains bag boundary conditions with opposite orientations of the normal vector on each side.

We note that the rather formal derivation we considered, namely, getting the 
matching conditions from the integration of Dirac's equation, may 
be related to the more exhaustive study of different singular potentials for 
the Dirac equation in $1+1$ dimensions, considered in~\cite{Sundberg:2003tc}. 
Indeed, the transition matrix ${\mathbb T}$ from the left to the right of 
square pulse potential centered on $x^d = 0$, which tends to a  $\delta$ 
function of height $\lambda$ when its width tends to $0$ is:
\begin{equation}
	{\mathbb T} \;=\; e^{-\lambda} \, \mathcal{P}_+ + e^{\lambda} \, 
	\mathcal{P}_-
\end{equation}
which yields the same conditions we derived, under the correspondence: 
\begin{equation}
	e^{\lambda}\;=\; \frac{1 + g/2}{1 - g/2} \;.
\end{equation}

In most of this paper, we shall use the special cases corresponding to 
bag boundary conditions, where $|g_{L,R}| = 2$. So we will set 
$g_{L,R} = 2 \eta_{L,R}$,  with $\eta_{L,R} = \pm 1$, since the
sign choice affects the physics. However, in the next Section,
we present a more general result, valid for any real values of the 
constants. 

%====================================================================
%%%%%%%%%%%%%%%%%%%%%%%%%%%%%%%%%%%%%%%%%%%%%%%%%%%%%%%%%%%%%%%%%%%%%
%%%%%%%%%%%%%%%%%%%%%%%%%%%%%%%%%%%%%%%%%%%%%%%%%%%%%%%%%%%%%%%%%%%%%
%%%%%%%%%%%%%%%%%%%%%%%%%%%%% Casimir energy %%%%%%%%%%%%%%%%%%%%%%%%
%%%%%%%%%%%%%%%%%%%%%%%%%%%%%%%%%%%%%%%%%%%%%%%%%%%%%%%%%%%%%%%%%%%%%
%%%%%%%%%%%%%%%%%%%%%%%%%%%%%%%%%%%%%%%%%%%%%%%%%%%%%%%%%%%%%%%%%%%%%
\section{Casimir energy}\label{sec:Casimir}
One of the most immediate consequences of the boundary conditions imposed 
is in the vacuum energy which, since it depends on the distance $a$ between the 
walls, leads to Casimir forces. The vacuum energy of the confined Dirac field 
can be computed via the functional determinant of the operator appearing in the 
action~(\ref{eq:action}), with the singular potential inducing non-trivial 
boundary conditions at the walls. The result, after renormalization, depends on 
the separation $a$ between the walls and on the signs $\eta_{L,R}$.

For the case of two walls located at $x^d=a_L$ and $x^d=a_R\equiv a_L+a$, with 
coupling constants $g_L$ and $g_R$, the 
Casimir energy per unit area,  $\mathcal{E}(a)$,  may be obtained as follows: 
we first define
\begin{equation}\label{eq:defgammav}
e^{i \Gamma(V)}\;=\; \frac{\mathcal Z(V)}{\mathcal Z_0}
\end{equation}
with
\begin{equation}\label{eq:defzv}
	{\mathcal Z}(V)\;=\;\int {\mathcal D}\bar\psi {\mathcal D}\psi \,
	e^{i {\mathcal S}(\bar\psi,\psi; V)}\;\;{\rm and}\;\;\; {\mathcal 
	Z}_0\;\equiv\;	{\mathcal Z}(V)\big|_{V = 0} \;.
\end{equation}
Since the result of integrating out the fields may be formally put in terms of 
functional determinants:
\begin{equation}
\Gamma(V)\;=\; -i \, \log\Big[\frac{\det( i \not \!\partial \,-\, V)}{\det(i 
\not \!\partial)}\Big]
\end{equation}
Since $V(x^d)$ depends only on the perpendicular coordinate, the system is 
translation-invariant along the $d$ parallel space-time directions, so that the 
effective action factors as $\Gamma(V) = -\mathcal{E}(a)\,V_\shortparallel$, with 
$V_\shortparallel$ the (large) parallel space-time volume and $\mathcal{E}(a)$ 
the energy per unit parallel area. Fourier-decomposing the determinant along 
these directions, and dropping an $a$-independent constant (which accounts for 
the self-energy of the walls), we get
\begin{align}\label{eq:funct}
\mathcal{E}(a) \;=\; & i 
\int \frac{d^dp_\shortparallel}{(2\pi)^d} \log\Big[\frac{\det( i \gamma^d 
\partial_d + \not \! p_\shortparallel \,-\, V)}{\det( i \gamma^d 
\partial_d + \not \! p_\shortparallel)}\Big] \nonumber\\
\;=\;& i 
\int \frac{d^dp_\shortparallel}{(2\pi)^d} {\rm Tr}\log\Big[ \big({i \gamma^d 
	\partial_d + \not \! p_\shortparallel}\big)^{-1} \big( i \gamma^d 
	\partial_d + \not \! p_\shortparallel \,-\, V \big)\Big]\;.
\end{align}
Subtracting from $\mathcal{E}(a)$ its limit when $a \to \infty$ leads to the 
interaction Casimir energy. Since $V$ has support on a finite set of points, 
the Fredholm determinant in (\ref{eq:funct}) collapses to that of a spinor 
wall-index matrix already (see Appendix A). 

For arbitrary couplings $g_L, g_R$:
\begin{equation}\label{eq:E-detM-body}
\mathcal{E}(a) \;=\; i\int\frac{d^dp_\shortparallel}{(2\pi)^d}\,
\log\det M(p_\shortparallel),
\end{equation}
with $M(p_\shortparallel)$ given explicitly in 
Appendix~\ref{app:energy}. Its determinant evaluates, in any number of 
dimensions, to
\begin{equation}\label{eq:detM-body}
\det M(p_\shortparallel)\;=\;
\left[\frac{(4+g_L^2)(4+g_R^2)}{16}\right]^{2^{\lfloor n/2\rfloor-1}}
\!\bigl[1+\rho_L\rho_R\,e^{2iaP}\bigr]^{2^{\lfloor n/2\rfloor-1}},
\end{equation}
with $P\equiv\sqrt{p_\shortparallel^2}$ and the ``reflection coefficients''
\begin{equation}\label{eq:rho-body}
\rho_i\;\equiv\;\frac{4\,g_i}{4+g_i^2},\qquad i=L,R\,.
\end{equation}
Subtracting an $a$-independent term, and performing the Wick rotation, the 
renormalised Casimir energy  becomes
\begin{equation}\label{eq:E-arb-body}
\mathcal{E}(a)\;=\;-\,2^{\lfloor n/2\rfloor-1}\!\int\!
\frac{d^dp_\shortparallel}{(2\pi)^d}\,
\log\!\bigl[1+\rho_L\rho_R\,e^{-2a|p_\shortparallel|}\bigr]\,.
\end{equation}
The reflection coefficients satisfy $|\rho_i|\le 1$, with the bound 
saturated only when $|g_i|=2$; in that limit $\rho_i\to\eta_i$ and 
(\ref{eq:E-arb-body}) reduces to
\begin{equation}\label{eq:E-bag-body}
\mathcal{E}(a)\;=\;-\,2^{\lfloor n/2\rfloor-1}\!\int\!
\frac{d^dp_\shortparallel}{(2\pi)^d}\,
\log\!\bigl[1+\eta_L\eta_R\,e^{-2a|p_\shortparallel|}\bigr]\,.
\end{equation}
Equation~(\ref{eq:E-bag-body}) is therefore the perfect-confinement limit 
of the more general expression. For $|g_i|<2$ the magnitude of the Casimir 
energy is reduced, while its sign is still determined by 
$\mathrm{sgn}(\rho_L\rho_R)=\mathrm{sgn}(g_L g_R)$. In particular, for 
symmetric couplings $g_L=g_R=g$, $|\mathcal{E}(a)|$ reaches its maximum 
precisely at $|g|=2$; this was shown by explicit evaluation in $1+1$ and 
$3+1$ dimensions in~\cite{FoscoLosada2008}, and is consistent with 
(\ref{eq:E-arb-body}).

\medskip

Throughout the rest of the paper we work exclusively in the bag-condition 
limit $|g_{L,R}|=2$, where (\ref{eq:E-bag-body}) means that the force is 
attractive when $\eta_L=\eta_R$ and repulsive when $\eta_L=-\eta_R$. The 
repulsive case is consistent with the theorem established 
in~\cite{Fernandez:2025fpf}, which states that a sufficient condition for 
repulsion is that the potential coupling the fermionic field to the objects 
be odd under spatial reflection. A further reason for this restriction is 
that the surface-current correlators of Sec.~\ref{sec:correlation} and the 
induced bulk current of Sec.~\ref{sec:induced} depend on the full propagator 
within the slab, which factorises only at $|g_{L,R}|=2$ (cf.~Appendix~A).

Moreover, in this case the Casimir energy is exactly $2^{\lfloor\frac{n}{2}\rfloor}$ times the Casimir energy of a scalar field obeying Dirichlet boundary conditions, but with opposite sign (repulsive rather than attractive). 

A simple way to understand this result is that the odd version of the potential effectively imposes Dirichlet conditions on half of the $2^{\lfloor\frac{n}{2}\rfloor}$ spinor components on the inner sides of the walls, while leaving the remaining components unconstrained there. This ``half'' factor is compensated by the factor of $2$ appearing in the power of the determinant in the fermionic functional integral relative to the bosonic counterpart.

\subsection{The pressure in terms of a bilinear concentrated on one of the 
walls}
Let us now explore the relationship between the phenomenon of Casimir 
interaction and the vacuum expectation values of fermionic bilinears 
concentrated on one of the planes. We know that this must 
indeed be possible, since one could calculate a force by 
evaluating the corresponding flux of the energy-momentum tensor.
But we use here a more expedite procedure, namely, we derive the Casimir force 
per unit area, or pressure ${\mathcal P}(a)$, from~(\ref{eq:funct}). Indeed, 
applying the same kind of procedure as in \cite{Fosco2023}, we derive:
\begin{equation}\label{eq:pressure}
{\mathcal P}(a) = - \partial_a \mathcal{E}(a) \,=\,
- i\, g_R \, \int \frac{d^dp_\shortparallel}{(2\pi)^d} 	\,
\tr \Big[ \gamma^d \not \! p_\shortparallel 
\widetilde{S}_F(p_\shortparallel;a_R,a_R)\Big] 
\end{equation}
where $\widetilde{S}_F(p_\shortparallel;x^d,y^d)$ is the fermionic propagator 
in the presence of two walls, derived in Appendix~\ref{app:propagator}. This already shows that 
the pressure depends on a fermionic bilinear, concentrated on one of the 
plates. Of course, the opposite result would have been obtained had we 
considered the derivative with respect to $a_L$ instead.
In the particular case of $n=2+1$ dimensions, since $\gamma^d \gamma^\alpha = 
- i \varepsilon^{\alpha \beta} \gamma_\beta$, we can even refine the previous 
assertion to saying that the pressure depends on the fermionic current on the 
wall.

\medskip

A remark on the meaning of the propagator at the wall is in order. The 
propagator obtained in Appendix~\ref{app:propagator} is, in general, 
discontinuous across each wall: it contains a piece carrying a 
$\mathrm{sgn}(x^d-y^d)$ factor, reflecting the well-known Šeba-type jump in 
the Dirac wavefunction induced by a $\delta$-potential~\cite{Seba1989}. The 
trace appearing in~(\ref{eq:pressure}), however, projects out exactly the 
discontinuous components: using the decomposition of the propagator 
$\widetilde{S}_F = A\,\mathbb{1}+B\,\hat{p}_\alpha\gamma^\alpha+C\,\gamma^d
+D\,\hat{p}_\alpha\gamma^\alpha\gamma^d$ given in Appendix~\ref{app:propagator}, 
together with the Dirac algebra (with $g^{dd}=-1$), one finds in any 
dimension
\begin{equation}\label{eq:trace-id}
\tr\!\left[\gamma^d\not\!p_\shortparallel\,\widetilde{S}_F(p_\shortparallel;
x^d,y^d)\right]\;=\;-\,2^{\lfloor n/2\rfloor}\,D(p_\shortparallel;x^d,y^d)\,
\sqrt{p_\shortparallel^2}.
\end{equation}
Of the four scalar coefficients, only the smooth coefficient $D$ contributes 
to the pressure. $D$ at coincident points on the wall is, in turn, continuous. 
From the explicit expressions in 
Appendix~\ref{app:propagator},
\begin{equation}\label{eq:Dwall}
D(p_\shortparallel;a_R,a_R)\;=\;
\begin{cases}
(\eta/2)\,\tan(\sqrt{p_\shortparallel^2}\,a) & \text{(symmetric case)}\\[2pt]
(\eta/2)\,\cot(\sqrt{p_\shortparallel^2}\,a) & \text{(antisymmetric case)}
\end{cases}
\end{equation}
neither of which involves any sign function. The choice of side of the wall 
at which to evaluate the propagator is also unambiguous: for $|g_{L,R}|=2$ 
the wall is impermeable, and the pressure exerted on the right wall is 
determined by the propagator on its inner face, the only one in causal 
contact with the slab interior. This realises the partial cancellation 
expected on the grounds that, in the same sense in which a $\delta$-potential 
acts on a Dirac field as a $\delta'$-potential would on a 
scalar~\cite{Seba1989,Sundberg:2003tc}, the projected combination relevant 
for the pressure is well defined.

%====================================================================
%%%%%%%%%%%%%%%%%%%%%%%%%%%%%%%%%%%%%%%%%%%%%%%%%%%%%%%%%%%%%%%%%%%%%
%%%%%%%%%%%%%%%%%%%%%%%%%%%%%%%%%%%%%%%%%%%%%%%%%%%%%%%%%%%%%%%%%%%%%
%%%%%%%%%%%%%%%%%%%%%%%%% Correlation %%%%%%%%%%%%%%%%%%%%%%%%%%%%%%%
%%%%%%%%%%%%%%%%%%%%%%%%%%%%%%%%%%%%%%%%%%%%%%%%%%%%%%%%%%%%%%%%%%%%%
%%%%%%%%%%%%%%%%%%%%%%%%%%%%%%%%%%%%%%%%%%%%%%%%%%%%%%%%%%%%%%%%%%%%%
\section{Correlation between currents}\label{sec:correlation}
As we have just discussed, the Casimir pressure may be expressed 
in terms of the vacuum expectation value of a surface current.
Let us now see that the correlation between the two fermionic currents, each 
one concentrated on a wall, also plays a role in the Casimir interaction.
 Indeed, by following a procedure which is entirely analogous 
to the one of \cite{Fosco2023}, but with time-independent deformations, we 
now assume a small deformation  of each boundary, 
in such a way that $L$ and $R$ correspond to $x^d = - \Big(\frac{a}{2} 
+\xi_L({\vec x}_\shortparallel) \Big)$ and $x^d = + \Big(\frac{a}{2} 
+\xi_R({\vec x}_\shortparallel) \Big)$, respectively. Here, 
${\vec x}_\shortparallel$ are the first $d-1$ spatial coordinates. Besides, 
we assume that the deformations have zero spatial average. Were such an average 
different from zero, it could be removed by a redefinition of $a$.

Then, from (\ref{eq:defgammav}) and (\ref{eq:defzv}), we obtain for the lowest 
order correction to the vacuum energy involving both deformations, 
$E_{LR}$:
\begin{equation}
E_{LR}\,=\, \int d^d{\vec x}_\shortparallel \int d^d{\vec 
y}_\shortparallel 
\,\xi_L({\vec x}_\shortparallel) \, 
	\gamma({\vec x}_\shortparallel - {\vec y}_\shortparallel )
\,\xi_R({\vec y}_\shortparallel)	
	\;, 
\end{equation}
where $\gamma({\vec x}_\shortparallel) = \int 
\frac{d^{d-1}{\vec k}_\shortparallel}{(2\pi)^{d-1}}
e^{i {\vec k}_\shortparallel \cdot {\vec x}_\shortparallel} 
\int \frac{dk^0}{2\pi} \widetilde{\gamma}(k^0,{\vec k}_\shortparallel)$
and:
\begin{equation}
\widetilde{\gamma}( k_\shortparallel)\,=\, 4 i \eta_L \eta_R \, 
\int \frac{d^dp_\shortparallel}{(2\pi)^d} \, {\rm tr} \Big[\gamma^d 
\tilde{S}_F\left(p_\shortparallel + k_\shortparallel;-\frac{a}{2},\frac{a}{2}\right)
(\not \! p_\shortparallel + \not \! k_\shortparallel)
\gamma^d \not \! p_\shortparallel
\tilde{S}_F\left(p_\shortparallel ;\frac{a}{2},-\frac{a}{2}\right)
\Big] \;.
\end{equation}

For $d=2$, the energy reduces to
\begin{equation}
	E_{LR}\,=\,  4 i \eta_L 
	\eta_R \,\int dx^1 \int dy^1
	\,A^0_L(x^1) \, 
	\Pi_{00}(x^1 - y^1 )
	\,
	A_R^0(y^1)	
	\;, 
\end{equation}
where we introduced the vacuum polarization tensor:
\begin{equation}
	\widetilde{\Pi}_{\alpha\beta}( k_\shortparallel)\,=\, 
	\int \frac{d^dp_\shortparallel}{(2\pi)^d} \, {\rm tr} \Big[\gamma_\alpha
	\tilde{S}_F\left(p_\shortparallel + k_\shortparallel;-\frac{a}{2},\frac{a}{2}\right)
	\gamma_\beta 
	\tilde{S}_F\left(p_\shortparallel ;\frac{a}{2},-\frac{a}{2}\right)
	\Big] \;,
\end{equation}
and $A^0 = \partial_1 \xi$. Note that in this static case, the geometric gauge 
field has only temporal component. Besides, it vanishes when the deformation is 
constant. A constant contribution would in turn amount to a 
change in $a$, and therefore cannot appear as a deformation.
Thus, the interaction between the deformations depends on the correlation 
between the surface currents on the two walls: a vacuum 
polarization tensor, which will also play an important role 
in Sect.~\ref{sec:induced}.

We then focus on the correlation between currents at the walls 
($x^d=-\frac{a}{2}$ and $y^d=\frac{a}{2}$):
\begin{align}
\mathcal{C}^{\alpha\beta}(x_\shortparallel-y_\shortparallel)
&\equiv\expval{J^\alpha\left(x_\shortparallel,-\frac{a}{2}\right) 
J^\beta\left(y_\shortparallel,\frac{a}{2}\right)}=\notag\\
&=-\int\limits_{\slashed{p}_\shortparallel\slashed{k}_\shortparallel}
e^{-i(p_\shortparallel-k_\shortparallel)\cdot(x_\shortparallel-y_\shortparallel)}
\tr\left[\gamma^\alpha\tilde{S}_F\left(p_\shortparallel;-\frac{a}{2},\frac{a}{2}\right)
\gamma^\beta\tilde{S}_F\left(k_\shortparallel;\frac{a}{2},-\frac{a}{2}\right)\right]
\end{align}

We will mostly consider the equal-time correlator, i.e. we set $x_\shortparallel-y_\shortparallel=(0,\vec{x}_\shortparallel)$, which isolates the spatial dependence induced by the boundaries and avoids mixing it with retardation effects. Other time orderings can be obtained by keeping the full $x_\shortparallel-y_\shortparallel$ dependence and performing the corresponding analytic continuation.

\begin{figure}
\centering
\begin{tikzpicture}
\begin{axis}[
    width=12cm,
    height=7cm,
    grid=both,
    xmin=-3.0, xmax=3.0,
    xlabel={$x/a$},
    ylabel={$(2\pi a^2)^2\expval{J^1(0,-\frac{a}{2})J^1(x,\frac{a}{2})}$},
    legend pos=north east,
]
\addplot[blue, line width = 2 pt] table {C1121.dat};
\addlegendentry{$\eta_L=\eta_R$}

\addplot[orange, line width = 2 pt] table {C112anti.dat};
\addlegendentry{$\eta_L=-\eta_R$}
\end{axis}
\end{tikzpicture}
\caption{Position dependence of correlation between parallel current densities at the walls in $d=2$.}
\label{fig:c2}
\end{figure}
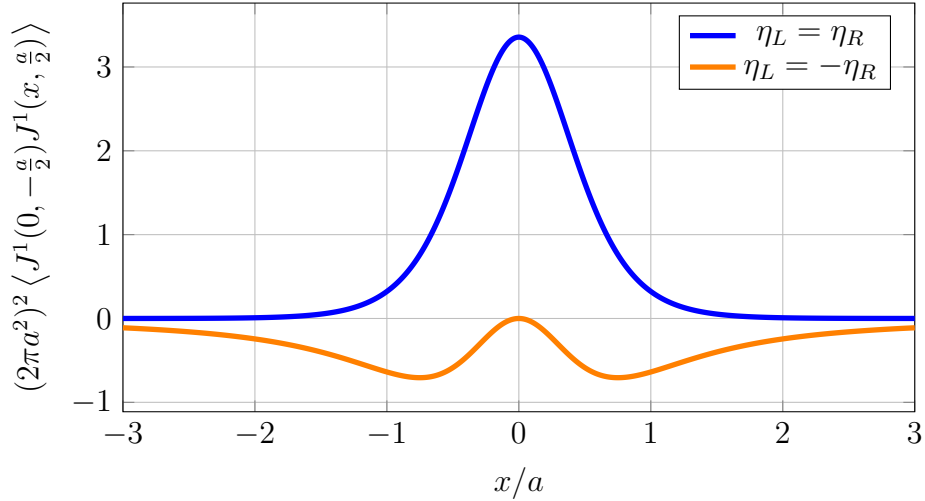

Using the expressions for the propagator from Appendix \ref{app:propagator} and performing the Dirac traces\footnote{Using that, for $n=2+1$ dimensions the gamma matrices can be chosen as $\gamma^\mu=(\sigma_1,i\sigma_2,i\sigma_3)$, we have $\gamma^\mu\gamma^\nu=g^{\mu\nu}\mathbb{1}-i\epsilon^{\mu\nu\rho}\gamma_\rho$ for $\mu,\nu,\rho\in\{0,1,2\}$. This gives $\tr(\gamma^\alpha\gamma^\beta\gamma^2)=-2i\epsilon^{\alpha\beta2}$ and $\tr(\gamma^\alpha\gamma^\rho\gamma^\beta\gamma^\lambda\gamma^2)=-2i(g^{\alpha\rho}\epsilon^{\beta\lambda2}+g^{\beta\lambda}\epsilon^{\alpha\rho2})$ for $\alpha,\beta,\rho,\lambda\in\{0,1\}$.} the equal-time correlator between currents at the walls is
\begin{align}
    \mathcal{C}^{\alpha\beta}(\vec{x}_\shortparallel)=-\frac{2^{\lfloor\frac{d-1}{2}\rfloor}}{(2\pi a^2)^d}\left(\frac{|\vec{x}_\shortparallel|}{a}\right)^{2-d}&\left\{ \delta_{\eta_L\eta_R}~\mathcal{U}^{\alpha\beta}~\mathcal{I}_d\left[\cosh(\cdot),\frac{|\vec{x_\shortparallel}|}{a}\right]+\right.\notag\\
    &\left.+\bar{\delta}_{\eta_L\eta_R}~\mathcal{V}^{\alpha\beta}(\vec{x}_\shortparallel)~~ \mathcal{I}_{d+2}\left[\cdot\sinh(\cdot),\frac{|\vec{x_\shortparallel}|}{a}\right]\right\},
\end{align}
where $\delta_{\eta_L\eta_R}$ is equal to one when $\eta_L=\eta_R$, zero otherwise, and $\bar\delta_{\eta_L\eta_R}$ the opposite. We have also defined
\begin{equation}
    \mathcal{I}_d\left[F(\cdot);t\right]=\left(\int\limits_0^\infty ds\frac{s^{d/2}J_{\frac{d-2}{2}}\left(s~t\right)}{F(s)}\right)^2,
\end{equation}
and the tensorial structure is given by
\begin{align}
    &\mathcal{U}^{\alpha\beta}=g^{\alpha\beta}-\delta_{d2}\eta_L\epsilon^{\alpha\beta 2}\\
    &\mathcal{V}^{\alpha\beta}(x)=g^{\alpha\beta}+2\frac{x^\alpha x^\beta}{-x^2}-\delta_{d2}\eta_L\frac{x_\rho x_\lambda}{-x^2}(g^{\alpha\rho}\epsilon^{\beta\lambda2}+g^{\beta\rho}\epsilon^{\alpha\lambda 2}).
\end{align}
The correlation between charge densities is always negative: a positive fluctuation at one wall tends to be accompanied by a negative fluctuation at the other. This reflects the tendency of the vacuum to balance charge fluctuations across the boundaries. Regarding the tensorial structure, a parity-breaking term arises in two spatial dimensions, and the antisymmetric case introduces an angular dependence. Let us analyze separately the most relevant cases of $d=2$ and $d=3$:

In two spatial dimensions with $x_\shortparallel=(0,x)$ the correlation between currents is
\begin{align}
    \mathcal{C}^{\alpha\beta}(x)=-\frac{1}{(2\pi a^2)^2}&\left\{\delta_{\eta_L\eta_R}\begin{pmatrix}
        1&-\eta_L\\\eta_L&-1
    \end{pmatrix}\left(\int_0^\infty \!\!\! ds\,\,\frac{s J_0\left(\frac{sx}{a}\right)}{\cosh(s)}\right)^2+\right.\\
    &\left.+\bar\delta_{\eta_L\eta_R}\begin{pmatrix}
        1&\eta_L\\\eta_L&1
    \end{pmatrix}\left(\int_0^\infty\!\!\! ds\,\,\frac{s J_1\left(\frac{sx}{a}\right)}{\sinh(s)}\right)^2\right\},
\end{align}
and the behavior of the integrals can be seen in Fig. \ref{fig:c2}, where the spatial dependence of the current-current correlation is shown. One observes that an even potential leads to a positively correlated response, whereas an odd configuration produces a negative correlation. This is consistent with the sign of the Casimir force: in analogy with Ampère's law, parallel (antiparallel) currents are associated with attractive (repulsive) interactions.

In three spatial dimensions the integrals can be explicitly computed, so the 
correlation between currents is, with $x_\shortparallel=(0,x,y)$ denoting the 
difference between the coordinates on each plane:   
\begin{align}
    \mathcal{C}^{\alpha\beta}(x,y)=-\frac{2}{(2\pi a^2)^3}\frac{a}{r}&\left\{\delta_{\eta_L\eta_R}\begin{pmatrix}
        1&0&0\\0&-1&0\\0&0&-1
    \end{pmatrix}\frac{\pi^3a}{8r}\frac{\tanh^2\left(\frac{\pi r}{2a}\right)}{\cosh^2\left(\frac{\pi r}{2a}\right)}\right.\\
    &\left.+\bar\delta_{\eta_L\eta_R}\begin{pmatrix}
        1&0&0\\0&\frac{x^2-y^2}{r^2}&\frac{2xy}{r^2}\\
        0&\frac{2xy}{r^2}&\frac{y^2-x^2}{r^2}
    \end{pmatrix}\frac{\pi a^3}{2r^3}\frac{\left(\sinh\left(\frac{\pi r}{a}\right)-\frac{\pi r}{a}\right)^2}{\left(1+\cosh\left(\frac{\pi r}{a}\right)\right)^2}\right\}\eval_{r=\sqrt{x^2+y^2}}\notag.
\end{align}
Note that, when the potential is  odd, there is an interesting phenomenon 
involving the spatial dependence of the correlation function. 
To that end, it is convenient to choose an appropriate coordinate system. Since 
$(x,y)$ and $(-y,x)$ are eigenvectors of the non-diagonal block in the 
correlation matrix, introducing polar coordinates on the plane: $\hat{r} = 
\frac{1}{r} (x,y)$ and $\hat{\phi} = \frac{1}{r} (-y, x)$ we see that
\begin{equation}
M=
\begin{pmatrix}
	1 & 0 & 0\\[4pt]
	0 & \dfrac{x^2-y^2}{r^2} & \dfrac{2xy}{r^2}\\[8pt]
	0 & \dfrac{2xy}{r^2} & \dfrac{y^2-x^2}{r^2}
\end{pmatrix}
\end{equation}
becomes diagonal:
\begin{equation}
	M'=
	\begin{pmatrix}
		1 & 0 & 0\\[4pt]
		0 & 1 & 0 \\[8pt]
		0 & 0 & -1
	\end{pmatrix}
	\;.
\end{equation}
This means that the radial components and the tangential components of the 
currents are separetely correlated. Namely, in the odd case, the only 
non-vanishing correlations are:
\begin{equation}
\mathcal{C}_{rr}^{\mathrm odd}(x,y)  =-\mathcal{C}_{\phi \phi}^{\mathrm odd}(x,y)= -\frac{\pi}{(2\pi a^2)^3}
\big(\frac{a}{r}\big)^4  \Big[ \frac{\sinh(\frac{\pi r}{a}) - 
\frac{\pi r}{a} }{ 1 + \cosh(\frac{\pi r}{a}) } \Big]^2 \;.
\end{equation}
One way to have a pictorial representation of the previous result is 
that fluctuations that look like a radial flow from (or towards) a point
on a plane are negatively correlated. On the other hand, fluctuations looking 
like circular streams have a positive correlation.

%====================================================================
%%%%%%%%%%%%%%%%%%%%%%%%%%%%%%%%%%%%%%%%%%%%%%%%%%%%%%%%%%%%%%%%%%%%%
%%%%%%%%%%%%%%%%%%%%%%%%%%%%%%%%%%%%%%%%%%%%%%%%%%%%%%%%%%%%%%%%%%%%%
%%%%%%%%%%%%%%%%%%%%%%%%%%%%% Induced %%%%%%%%%%%%%%%%%%%%%%%%%%%%%%%%%
%%%%%%%%%%%%%%%%%%%%%%%%%%%%%%%%%%%%%%%%%%%%%%%%%%%%%%%%%%%%%%%%%%%%%
%%%%%%%%%%%%%%%%%%%%%%%%%%%%%%%%%%%%%%%%%%%%%%%%%%%%%%%%%%%%%%%%%%%%%
\section{Induced bulk current}\label{sec:induced}

\begin{figure}
\begin{tikzpicture}
\begin{axis}[
    width=12cm,
    height=7cm,
    grid=both,
    xmin=-0.5, xmax=0.5,
    xlabel={$x/a$},
    ylabel={$\frac{4\pi}{e^2E}j^1(x)$}, y label style={rotate=-90},
    legend pos=north east,
]
\addplot[blue, line width = 2 pt] table {jsim.dat};
\addlegendentry{$\eta_L=\eta_R$}

\addplot[orange, line width = 2 pt] table {janti.dat};\draw[black,-stealth, line width = 2 pt]
  (rel axis cs:0.35,0.25) -- (rel axis cs:0.65,0.25)
  node[midway, above] {$E$};
\addlegendentry{$\eta_L=-\eta_R$}
\end{axis}
\end{tikzpicture}
\caption{Spatial current density in the presence of an external electric field perpendicular to the walls.}
\label{fig:j}
\end{figure}
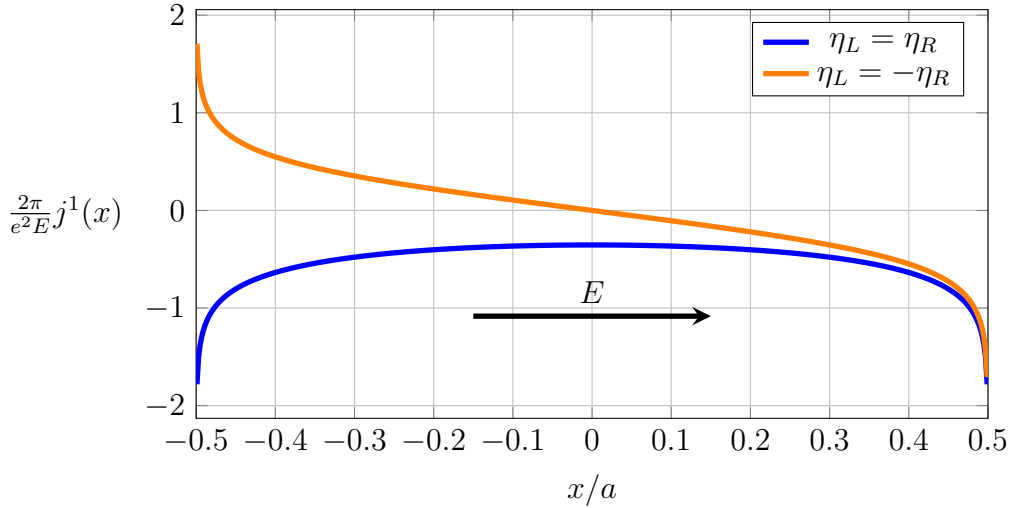

The induced current is defined as the vacuum expectation value of the fermionic current $e\,\bar\psi(x)\gamma^\mu\psi(x)$ (with $e<0$ the charge of the electron) in the presence of an external electromagnetic field $A_\mu$. We are interested in the lowest order contribution, i.e. the linear response, which turns out to be
\begin{equation}
    j^\mu(x)=\expval{ e\bar\psi(x)\gamma^\mu\psi(x)}_A\eval_{1\text{st order}}\hspace{-5mm}=ie^2\int d^ny~\Pi^{\mu\nu}(x,y)A_\nu(y),
\end{equation}
where $\Pi^{\mu\nu}(x,y)=\tr\left(\gamma^\mu S_F(x,y)\gamma^\nu S_F(y,x)\right)$ is the vacuum polarization tensor. For a uniform electric field perpendicular to the walls $\vec{E}=(\vec{0}_\shortparallel,E)$, with the gauge choice $A_\mu(x)=(-x^dE,\vec{0})$, the induced current takes the form
\begin{equation}\label{eq:jmu}
    j^\mu(x^d)=-ie^2E\int\limits_{-\infty}^\infty dy^d~y^d\tilde{\Pi}^{\mu0}(0_\shortparallel;x^d,y^d),
\end{equation}
with
\begin{equation}
    \tilde{\Pi}^{\mu0}(0_\shortparallel;x^d,y^d)=\int\frac{d^dp_\shortparallel}{(2\pi)^d}\tr[\gamma^\mu\tilde{S}_F(p_\shortparallel;x^d,y^d)\gamma^0\tilde{S}_F(p_\shortparallel;y^d,x^d)].
\end{equation}
If we are only interested in what happens between the walls, the limits of 
(\ref{eq:jmu}) can be replaced by $\left(-\frac{a}{2},\frac{a}{2}\right)$. 
Taking into account the symmetry properties of the coefficients found in 
Appendix~\ref{app:propagator}, after performing the Dirac traces\footnote{This expression is 
valid for $d=1,2,3$} we get
\begin{align}
    \tilde{\Pi}^{\mu 0}(0_\shortparallel;x^d,y^d)=\delta^{\mu0}\tr\mathbb{1}&\int\frac{d^dp_\shortparallel}{(2\pi)^d}\left[A^2-C^2+\frac{p_0^2+\vec{p}_\shortparallel^2}{p_0^2-\vec{p}_\shortparallel^2}(B^2-D^2)\right]+\\
    +4i\delta^{\mu 1}\delta^{d2}&\int\frac{d^dp_\shortparallel}{(2\pi)^d}\left[\frac{p_0^2+p_1^2}{p_0^2-p_1^2}BD-AC\right],\label{eq:corriente}
\end{align}
This expression shows that there is only induced spatial current for the case of $2+1$ dimensions, and it is transversal to the electric field. After the Wick rotation $p^0\to ip_E^0$, the first term in (\ref{eq:corriente}) --- the one involving the product $BD$ --- vanishes by angular integration in the Euclidean parallel-momentum plane, as we now show. The coefficients $B$ and $D$ depend on $p_\shortparallel$ only through $\sqrt{p_\shortparallel^2}$ (cf.~Appendix~A), and hence become functions of the Euclidean radial variable $p_E\equiv\sqrt{(p_E^0)^2+p_1^2}$ alone, with no dependence on the angular variable $\varphi$ defined by $(p_E^0,p_1)=p_E(\sin\varphi,\cos\varphi)$. The kinematic factor multiplying them, in turn, transforms as
\begin{equation}\label{eq:angular}
\frac{p_0^2+p_1^2}{p_0^2-p_1^2}\;\xrightarrow{\text{Wick}}\;
-\,\frac{p_1^2-(p_E^0)^2}{p_E^2}\;=\;-\cos(2\varphi),
\end{equation}
and $\int_0^{2\pi}\!\cos(2\varphi)\,d\varphi=0$, so its integration over the angular variable annihilates the $BD$ contribution. The $AC$ term carries no analogous angular factor (after Wick rotation, $A$ and $C$ are also functions of $p_E$ only) and is thus unaffected. The induced spatial current therefore reduces to
\begin{equation}
    j^1(x)\equiv\eta_R\frac{e^2E}{4\pi}~\tilde{j}_{\eta_L\eta_R}\left(\frac{x}{a}\right)=\eta_R\frac{e^2E}{4\pi}\int\limits_{-\frac{1}{2}}^\frac{1}{2}dt~t\int\limits_{0}^\infty ds~s~f\left(s;\frac{x}{a},t;\eta_L\eta_R\right),
\end{equation}
where now we denote $x$ to the coordinate transversal to the walls, and with
\begin{align}
    &f(s;z,t;+1)=2\frac{\cosh[s(z+t)]}{\cosh(s)}\left[\mathrm{sgn}(z-t)e^{-s|z-t|}+e^{-s}\frac{\sinh[s(z-t)]}{\cosh(s)}\right]\\
    &f(s;z,t;-1)=2\frac{\sinh[s(z+t)]}{\sinh(s)}\left[\mathrm{sgn}(z-t)e^{-s|z-t|}-e^{-s}\frac{\sinh[s(z-t)]}{\sinh(s)}\right].
\end{align}
Evaluating the integrals yields
\begin{align}
    \tilde{j}_{+1}(x)=&-\frac{1}{2}-\frac{1}{6}\log2+6\log A_\text{GK}+\frac{1}{2}\log\left[\frac{\Gamma\left(\frac{3-2x}{4}\right)\Gamma\left(\frac{3+2x}{4}\right)}{\Gamma\left(\frac{1-2x}{4}\right)\Gamma\left(\frac{1+2x}{4}\right)}\right]+\\
    &-\frac{x}{4}\left[\psi_0\left(\frac{1-2x}{4}\right)-\psi_0\left(\frac{1+2x}{4}\right)-\psi_0\left(\frac{3-2x}{4}\right)+\psi_0\left(\frac{3+2x}{4}\right)\right]+\notag\\
    &-\frac{1-4x^2}{64}\left[\psi_1\left(\frac{1-2x}{4}\right)+\psi_1\left(\frac{1+2x}{4}\right)-\psi_1\left(\frac{3-2x}{4}\right)-\psi_1\left(\frac{3+2x}{4}\right)\right],\notag\\
    \tilde{j}_{-1}(x)=&\frac{1}{2}\log\left[\frac{\Gamma\left(\frac{1+2x}{2}\right)}{\Gamma\left(\frac{1-2x}{2}\right)}\right]-\frac{x}{2}\left[\psi_0\left(\frac{1-2x}{2}\right)+\psi_0\left(\frac{1+2x}{2}\right)\right]+\\
    &-\frac{1-4x^2}{16}\left[\psi_1\left(\frac{1-2x}{2}\right)-\psi_1\left(\frac{1+2x}{2}\right)\right]\notag
\end{align}
where $A_{GK}\approx1.2824...$ is the Glaisher-Kinkelin constant, which arises 
from the regularization of the infinite product of $\Gamma$ functions, and 
$\psi_0,\psi_1$ are the polygamma functions of order $0$ and $1$ respectively. We 
can see in Fig.~\ref{fig:j} that an even confining potential produces an 
even current density, while an odd confining potential 
leads to an odd distribution. This means that an external electric field with 
an even confining potential produces a net induced current in the direction 
perpendicular to the electric field.

Regarding the charge density, one can verify that in both configurations the distribution is an odd function of $x^d$ for every dimension, so the total induced charge between the walls vanishes, as expected.
%\begin{figure}
%\begin{tikzpicture}
%\begin{axis}[
%    width=12cm,
%    height=7cm,
%    grid=both,
%    xmin=-0.5, xmax=0.5,
%    xlabel={$x/a$},
%    ylabel={$\frac{2\pi}{e^2E}j^1(x)$}, y label style={rotate=-90},
%]
%\addplot[blue, line width = 2 pt] table {jsim.dat};\draw[black,-stealth, line 
%width = 2 pt]
%  (rel axis cs:0.35,0.5) -- (rel axis cs:0.65,0.5)
%  node[midway, above] {$E$};
%\end{axis}
%\end{tikzpicture}
%\caption{Spatial current density for an even confining potential in the 
%presence of an external electric field perpendicular to the walls.}
%\label{fig:jsim}
%\end{figure}
%
%\begin{figure}
%\begin{tikzpicture}
%\begin{axis}[
%    width=12cm,
%    height=7cm,
%    grid=both,
%    xmin=-0.5, xmax=0.5,
%    xlabel={$x/a$},
%    ylabel={$\frac{2\pi}{e^2E}j^1(x)$}, y label style={rotate=-90},
%]
%\addplot[orange, line width = 2 pt] table {janti.dat};\draw[black,-stealth, 
%line width = 2 pt]
%  (rel axis cs:0.35,0.75) -- (rel axis cs:0.65,0.75)
%  node[midway, above] {$E$};
%\end{axis}
%\end{tikzpicture}
%\caption{Spatial current density for an odd confining potential in the 
%presence of an external electric field perpendicular to the walls.}
%\label{fig:jasim}
%\end{figure}

%\begin{figure}
%\centering
%\begin{tikzpicture}
%\begin{axis}[
%    width=10cm,
%    height=7cm,
%    xmin=-0.55, xmax=0.55,
%    ymin=-1.2, ymax=1.2,
%    xlabel={$x^d/a$},
%    ylabel={$x^1/a$},
%    xtick={-0.5,0,0.5},
%    ytick={-1,-0.5,0,0.5,1},
%    axis lines=middle,
%    axis line style={-stealth},
%    clip=false,
%]
%\addplot[blue,-stealth, line width=1.5pt, 
%    quiver={u=0, v=\thisrowno{3}*0.7, scale arrows=0.05},
%    ] table [x index=0, y index=1] {jsim_field.dat};
%\draw[dashed, thick, gray] (-0.5,-1.2) -- (-0.5,1.2);
%\draw[dashed, thick, gray] (0.5,-1.2) -- (0.5,1.2);
%\node[gray, anchor=south] at (-0.5,1.2) {$x^d=-\frac{a}{2}$};
%\node[gray, anchor=south] at (0.5,1.2) {$x^d=\frac{a}{2}$};
%\draw[red,-stealth, line width = 2 pt]
%  (0.1,-1) -- (0.4,-1)
%  node[midway, below] {$\vec{E}$};
%\end{axis}
%\end{tikzpicture}
%\caption{Vector field representation of the induced current for the symmetric 
%configuration ($\eta_L=\eta_R$). The current flows parallel to the walls and 
%perpendicular to the applied electric field, constituting a Hall-like 
%response. 
%The magnitude increases near the boundaries.}
%\label{fig:jfield}
%\end{figure}

\subsection{Effective Hall conductivity and the parity anomaly}
The results of the previous subsection admit a natural interpretation in the 
language of the parity anomaly in $2+1$ 
dimensions~\cite{Redlich:1983dv,Niemi:1983rq}. In this dimensionality, a
single massless Dirac fermion cannot be regularized in a way that preserves 
both gauge invariance and parity; the parity anomaly manifests through a 
half-integer contribution to the Hall conductivity.
For the contribution 
to the parity anomaly of the borders in a four dimensional theory, 
see~\cite{Vass}, where sign-dependent boundary conditions (analogue to our 
choice for $\eta_{L,R}$) also appear.

In our system, the boundary conditions imposed by the singular potential 
project out half of the spinor components at each wall. For the symmetric 
configuration ($\eta_L=\eta_R$), the same projector $\mathcal{P}_\pm=(1\pm i\gamma^d)/2$ is imposed at both boundaries, effectively breaking parity within the slab. This is the mechanism 
behind the nonvanishing transverse current: the fermionic vacuum in the slab 
develops a parity-odd response to the external electric field.

An effective Hall conductivity can be defined by integrating the current density 
across the slab:
\begin{equation}\label{eq:sigma_H}
    J^1 \;\equiv\; \frac{1}{a}\int_{-a/2}^{a/2} dx^d\; j^1(x^d) 
    \;=\; \eta_R\,\frac{e^2 E}{4\pi}\int_{-1/2}^{1/2}dx\;\tilde{j}_{\eta_L\eta_R}(x)
    \;\equiv\; \sigma_H\, E\;.
\end{equation}
For the antisymmetric case, the odd symmetry of $\tilde{j}_{-1}(z)$ under 
$z\to-z$ guarantees that $\sigma_H=0$, consistent with the fact that the odd 
potential preserves parity.

For the symmetric case, integrating the analytic expression for 
$\tilde{j}_1(z)$ yields
\begin{equation}\label{eq:sigma_num}
    \sigma_H \;=-\;\eta_R\frac{e^2}{4\pi} \;,
\end{equation}
which is the natural parity anomaly scale, and it results from the bulk modes 
inhabiting the slab. The propagator coefficients  in the symmetric 
configuration (Appendix~A) have poles at 
$\cos(\sqrt{p_\shortparallel^2}\,a)=0$, i.e., at $p_\shortparallel^2 = m_n^2$ 
with
\begin{equation}\label{eq:slab-spectrum}
m_n\;=\;\left(n+\tfrac{1}{2}\right)\frac{\pi}{a},\qquad n=0,1,2,\ldots,
\end{equation}
which are precisely the effective $(2+1)$-dimensional masses of the discrete 
tower of eigenmodes of the perpendicular Dirac operator under the symmetric 
bag boundary conditions. Each such mode, viewed in the parallel directions 
as an effective $(2+1)$-dimensional massive Dirac fermion, contributes its 
own parity-anomaly response to the Hall conductivity; with 
in~(\ref{eq:sigma_num}) resulting from the sum of these contributions, with the 
regularisation implemented implicitly by the propagator-based derivation 
through the Wick rotation and the standard $i\varepsilon$ prescription. 

In the antisymmetric configuration the 
analogous poles lie at $\sin(\sqrt{p_\shortparallel^2}\,a)=0$, giving the 
spectrum $m_n=n\pi/a$ ($n=1,2,\ldots$); the parity-conjugate boundary 
conditions of this case organise the modes in $\pm m_n$ pairs, whose 
individual parity-anomaly contributions cancel pairwise, consistent with 
$\sigma_H=0$.

Finally, we note that the divergence of the current density near the walls 
is of logarithmic type, similar to the behavior of the vacuum 
expectation value of the stress-energy tensor near boundaries with 
Dirichlet-type conditions.
%====================================================================
%%%%%%%%%%%%%%%%%%%%%%%%%%%%%%%%%%%%%%%%%%%%%%%%%%%%%%%%%%%%%%%%%%%%%
%%%%%%%%%%%%%%%%%%%%%%%%%%%%%%%%%%%%%%%%%%%%%%%%%%%%%%%%%%%%%%%%%%%%%
%%%%%%%%%%%%%%%%%%%%%%%%%%% Conclusions %%%%%%%%%%%%%%%%%%%%%%%%%%%%%
%%%%%%%%%%%%%%%%%%%%%%%%%%%%%%%%%%%%%%%%%%%%%%%%%%%%%%%%%%%%%%%%%%%%%
%%%%%%%%%%%%%%%%%%%%%%%%%%%%%%%%%%%%%%%%%%%%%%%%%%%%%%%%%%%%%%%%%%%%%
\section{Discussion and Conclusions}\label{sec:Conclusions}

We have studied the vacuum physics of a massless Dirac field confined between 
two parallel walls in $d+1$ dimensions, realized through singular potentials 
with coupling constants $g_{L,R}=\pm 2$ that implement MIT bag boundary 
conditions. The two physically inequivalent configurations, corresponding to a confining potential that is even ($\eta_L=\eta_R$) or odd ($\eta_L=-\eta_R$) under reflection about the midplane, control the qualitative behavior of all observables.

For the Casimir energy, the even configuration yields an attractive force 
proportional to $2^{\lfloor n/2\rfloor}$ times the scalar Dirichlet result, 
while the odd configuration produces a repulsive force of the same magnitude.

The vacuum current-current correlator between the walls, 
$\mathcal{C}^{\alpha\beta}$, displays a rich tensorial structure. In arbitrary 
dimension, a parity-even piece proportional to $\eta^{\alpha\beta}$ (or the more 
involved tensor $\mathcal{V}^{\alpha\beta}$) mediates the standard 
vacuum-dressed Coulomb and Amp\`ere interactions between charges and currents 
on the walls. In $2+1$ dimensions, a parity-odd term proportional to 
$\epsilon^{\alpha\beta 2}$ appears, reflecting the parity anomaly of a single 
Dirac fermion. We showed that this term generates an effective Chern-Simons-like 
coupling between probe gauge fields on the two walls, present only in the 
symmetric configuration.

The most striking observable consequence emerges in the induced current: an 
external electric field perpendicular to the walls drives a transverse (Hall) 
current that flows parallel to the walls and perpendicular to the applied field, 
exclusively in $2+1$ dimensions. For the symmetric configuration, the current 
density is an even function of position, yielding a net integrated Hall current 
characterized by an effective conductivity $\sigma_H 
\;=-\;\eta_R\frac{e^2}{4\pi}$. For the antisymmetric configuration, the current 
density is odd, 
giving vanishing total current in agreement with parity conservation.

Several directions merit further investigation. The inclusion of a fermion mass 
$m$ would introduce a gap and is expected to drive the Hall conductivity toward 
different quantization values as  $ma\to\infty$, interpolating between 
different topological regimes when $m$ is finite. The extension to finite 
temperature would allow 
contact 
with the thermal Casimir effect and the behavior of the Hall response in a 
finite-temperature Dirac system. Finally, the interplay between the 
vacuum-mediated inter-wall interaction and the dynamical electromagnetic field, including the backreaction on the photon propagator within the slab, 
could reveal nontrivial modifications to the effective low-energy theory of 
the gauge field in the presence of fermionic boundary conditions.
\section*{Acknowledgments}
The authors thank ANPCyT, CONICET and UNCuyo for financial support.

\newpage
%====================================================================
%%%%%%%%%%%%%%%%%%%%%%%%%%%%%%%%%%%%%%%%%%%%%%%%%%%%%%%%%%%%%%%%%%%%%
%%%%%%%%%%%%%%%%%%%%%%%%%%%%%%%%%%%%%%%%%%%%%%%%%%%%%%%%%%%%%%%%%%%%%
%%%%%%%%%%%%%%%%%%%%%%%%%%%%% Appendix %%%%%%%%%%%%%%%%%%%%%%%%%%%%%%
%%%%%%%%%%%%%%%%%%%%%%%%%%%%%%%%%%%%%%%%%%%%%%%%%%%%%%%%%%%%%%%%%%%%%
%%%%%%%%%%%%%%%%%%%%%%%%%%%%%%%%%%%%%%%%%%%%%%%%%%%%%%%%%%%%%%%%%%%%%
\appendix
\section{Fermion propagator}\label{app:propagator}
In this Appendix we present the explicit form of the Dirac propagator in the presence of one and two walls.

\medskip

For a matrix differential operator
\begin{align}
    \mathcal{O}(x,y)&=\mathcal{O}_0(x-y)-\mathcal{V}(x)\delta(x-y)\\&=\int_z\left[\delta(x-z)\mathbb{1}-\mathcal{V}(x)\mathcal{O}_0^{-1}(x-z)\right]\mathcal{O}_0(z-y)\\&\equiv\int_zK(x,z)\mathcal{O}_0(z-y),
\end{align}
the inverse of $\mathcal{O}(x,y)$ is given by
\begin{equation}
    \mathcal{O}^{-1}(x,y)=\int_z\mathcal{O}_0^{-1}(x-z)K^{-1}(z,y).    
\end{equation}

For the inverse of $K$ we have
\begin{align}
    K^{-1}&=\frac{1}{1-\mathcal{V}\mathcal{O}_0^{-1}}=\sum_{n=0}^\infty(\mathcal{V}\mathcal{O}_0^{-1})^n=1+\sum_{n=1}^\infty(\mathcal{V}\mathcal{O}_0^{-1})^n\\
    &=1+\mathcal{V}\mathcal{O}_0^{-1}\sum_{n=0}^\infty(\mathcal{V}\mathcal{O}_0^{-1})^{n}=1+\mathcal{V}\mathcal{O}_0^{-1}K^{-1}=1+\mathcal{V}\mathcal{O}^{-1}
\end{align}
so for the inverse of $\mathcal{O}$ we get the implicit equation
\begin{equation}
    \mathcal{O}^{-1}(x,y)=\mathcal{O}_0^{-1}(x-y)+\int_z\mathcal{O}_0^{-1}(x-z)\mathcal{V}(z)\mathcal{O}^{-1}(z,y).
\end{equation}

\medskip

The complete propagator is defined as the inverse of the kinetic operator (up to a factor $i$):
\begin{equation}
    S_F(x,y)\equiv i[i\gamma^\mu\partial_\mu-V]^{-1}(x,y),
\end{equation}
and, as shown in the previous step, it satisfies the implicit equation
\begin{equation}\label{eq:implicit}
    S_F(x,y)=S_F^{(0)}(x-y)-i\int d^n z S^{(0)}_F(x-z)V(z)S_F(z,y),
\end{equation}
where
\begin{equation}
   S^{(0)}_F(x-y)\equiv i[i\gamma^\mu\partial_\mu]^{-1}(x-y)=i\int\frac{d^np}{(2\pi)^n}\frac{ p_\mu\gamma^\mu}{p^2+i\varepsilon}e^{-ip\cdot(x-y)} 
\end{equation}
is the free propagator for a massless Dirac field. For the configuration considered in this paper, because of the translational symmetry in the directions parallel to the walls, the full propagator can be expressed as
\begin{equation}
    S_F(x,y)=\int\frac{d^dp_\shortparallel}{(2\pi)^d}e^{-ip_\shortparallel\cdot(x_\shortparallel-y_\shortparallel)}\tilde{S}_F(p_\shortparallel;x^d,y^d).
\end{equation} 
Inserting the potential (\ref{eq:defv}) in (\ref{eq:implicit}), the (partial) Fourier transform of the full propagator turns out to be
\begin{equation}\label{eq:eqPropagator}
    \tilde{S}_F(p_\shortparallel;x^d,y^d)=\tilde{S}_F^{(0)}(p_\shortparallel;x^d-y^d)-i\sum_{i,j=1}^Ng_i\tilde{S}^{(0)}_F(p_\shortparallel;x^d-a_i)\,T_{ij}\,\tilde{S}^{(0)}_F(p_\shortparallel;a_j-y^d),
\end{equation}
where the (partial) Fourier transform of the free propagator is\footnote{$p_\shortparallel^2$ is implicitly defined with the Feynman prescription, namely $p_\shortparallel^2\equiv p_\shortparallel^2+i\epsilon$ with $\epsilon\to0^+$.}
\begin{equation}
    \tilde{S}_F^{(0)}(p_\shortparallel;z)=\frac{e^{i|z|\sqrt{p_\shortparallel^2}}}{2}\left(\frac{p_\alpha}{\sqrt{p_\shortparallel^2}}\gamma^\alpha-\text{sgn}(z)\gamma^d\right),~~\text{with }\alpha=0,\dots,d-1
\end{equation}
and $T_{ij}$ are matrices in the spinor space satisfying \begin{equation}\label{eq:T}
    \sum_{k=1}^N T_{ik}\left[\delta_{kj}\mathbb{1}+ig_j\tilde{S}_F^{(0)}(p_\shortparallel;a_k-a_j)\right]=\delta_{ij}\mathbb{1}.
\end{equation}

\subsection{One wall}
For one wall placed at $x^d=a$ with arbitrary $g$, the propagator takes the 
following form:
\begin{itemize}
	\item When both points are evaluated at the same side of the wall, 
	satisfying $\text{sgn}(x^d-a)=\text{sgn}(y^d-a)$, the propagator is
	\begin{equation}
		\tilde{S}_F(p_\shortparallel;x^d,y^d)=\tilde{S}^{(0)}_F(p_\shortparallel;x^d-y^d)-i\,\text{sgn}(x^d-a)\frac{g}{1+\left(\frac{g}{2}\right)^2}\tilde{S}^{(0)}_F(p_\shortparallel;x^d+y^d-2a)\gamma^d.
	\end{equation}
	
	\item When each point is evaluated at a different side of the wall, so that 
	$\text{sgn}(x^d-a)=-\text{sgn}(y^d-a)$, the propagator is
	\begin{equation}
		\tilde{S}_F(p_\shortparallel;x^d,y^d)=\frac{1-\left(\frac{g}{2}\right)^2}{1+\left(\frac{g}{2}\right)^2}\tilde{S}^{(0)}_F(p_\shortparallel;x^d-y^d).
	\end{equation}
	For $g=\pm2$, the field cannot propagate from one side of the wall to the 
	other, so this is another way of seeing that the singular potential with 
	that particular choice of the parameter imposes perfect bag boundary 
	conditions.
\end{itemize}

\subsection{Two walls with $|g|=2$}

For two walls located at $x^d=a_L$ and $x^d=a_R>a_L$, with coupling constants 
$g_L=2\eta_L$ and $g_R=2\eta_R$ respectively, for $\eta_{L,R}=\pm1$, we need to find the inverse of a matrix whose components are given by the 
bracket of ~(\ref{eq:T}), which is
\begin{equation}
	\begin{pmatrix}
		\mathbb{1}+2i\eta_L\tilde{S}^{(0)}_F(p_\shortparallel;0)&2i\eta_R\tilde{S}^{(0)}_F(p_\shortparallel;a_L-a_R)
		 \\
		2i\eta_L\tilde{S}^{(0)}_F(p_\shortparallel;a_R-a_L)&\mathbb{1}+2i\eta_R\tilde{S}^{(0)}_F(p_\shortparallel;0)
	\end{pmatrix}
\end{equation}
Using the expression for the inverse of a block matrix, after some algebra the 
full propagator can be expressed as
\begin{equation}
	\tilde{S}_F(p_\shortparallel;x^d,y^d)=A\,\mathbb{1}+B\,\hat{p}_\alpha\gamma^\alpha+C\,\gamma^d+D\,\hat{p}_\alpha\gamma^\alpha\gamma^d,
\end{equation}
where $\hat{p}_\alpha\equiv\frac{p_\alpha}{\sqrt{p_\shortparallel^2}}$, and 
$A,B,C$ and $D$ are scalar functions that make the propagator satisfy the bag 
boundary conditions at both sides of each wall:
\begin{equation}
	\lim_{\epsilon\to0^+}\left(\mathbb{1}\pm 
	i\gamma^d\right)\tilde{S}_F(p_\shortparallel;a_i\mp\eta_i\epsilon,y^d)=0,~~i=L,R.
\end{equation}

For the sake of symmetry let us take $a_L=-\frac{a}{2}$ and $a_R=\frac{a}{2}$. 
Since we are interested in what happens between the walls, we will only give 
the expressions of the coefficients for 
$x^d,y^d\in\left[-\frac{a}{2},\frac{a}{2}\right]$:
\begin{itemize}
	\item Symmetric case $\eta_L=\eta_R\equiv\eta$:
	\begin{align}
&A(p_\shortparallel;x^d,y^d)=-\frac{i\eta}{2}
\frac{\cos\sqrt{p_\shortparallel^2}(x^d+y^d)}{\cos\sqrt{p_\shortparallel^2}a}\\
&B(p_\shortparallel;x^d,y^d)=\frac{e^{i\sqrt{p_\shortparallel^2}|x^d-y^d|}}{2}-
\frac{e^{i\sqrt{p_\shortparallel^2}a}}{2}\frac{\cos\sqrt{p_\shortparallel^2}
(x^d-y^d)}{\cos\sqrt{p_\shortparallel^2}a}\\
&C(p_\shortparallel;x^d,y^d)=-\mathrm{sgn}(x^d-y^d)
\frac{e^{i\sqrt{p_\shortparallel^2}|x^d-y^d|}}{2}+
i\frac{e^{i\sqrt{p_\shortparallel^2}a}}{2}
\frac{\sin\sqrt{p_\shortparallel^2}(x^d-y^d)}{\cos\sqrt{p_\shortparallel^2}a}\\
&D(p_\shortparallel;x^d,y^d)=\frac{\eta}{2}
\frac{\sin\sqrt{p_\shortparallel^2}(x^d+y^d)}{\cos\sqrt{p_\shortparallel^2}a}
\end{align}
	
\item Antisymmetric case $\eta_L=-\eta_R\equiv\eta$:
\begin{align}
&A(p_\shortparallel;x^d,y^d)=\frac{i\eta}{2}
\frac{\sin\sqrt{p_\shortparallel^2}(x^d+y^d)}{\sin\sqrt{p_\shortparallel^2}a}\\
&B(p_\shortparallel;x^d,y^d)=\frac{e^{i\sqrt{p_\shortparallel^2}|x^d-y^d|}}{2}
+i\frac{e^{i\sqrt{p_\shortparallel^2}a}}{2}\frac{\cos\sqrt{p_\shortparallel^2}(x^d-y^d)}{\sin\sqrt{p_\shortparallel^2}a}\\
&C(p_\shortparallel;x^d,y^d)=-\mathrm{sgn}(x^d-y^d)\frac{e^{i\sqrt{p_\shortparallel^2}|x^d-y^d|}}{2}+\frac{e^{i\sqrt{p_\shortparallel^2}a}}{2}\frac{\sin\sqrt{p_\shortparallel^2}(x^d-y^d)}{\sin\sqrt{p_\shortparallel^2}a}\\
&D(p_\shortparallel;x^d,y^d)=\frac{\eta}{2}\frac{\cos\sqrt{p_\shortparallel^2}(x^d+y^d)}{\sin\sqrt{p_\shortparallel^2}a}
\end{align}
\end{itemize}
It can be seen that for both cases $A,B$ and $D$ are even under the exchange 
$x^d\leftrightarrow y^d$, while $C$ is odd.

\medskip

Since a wall with $|g|=2$ completely decouples the two sides, when both points 
lie beyond all the walls the propagator reduces to that of a single wall:
\begin{equation}
	\tilde{S}_F(p_\shortparallel;x^d,y^d)=\begin{cases}
		\tilde{S}^{(0)}_F(p_\shortparallel;x^d-y^d)+i\eta_L\tilde{S}^{(0)}_F(p_\shortparallel;x^d+y^d-2a_L)\gamma^d&\text{if
		 } ~x^d,y^d<a_L\\
		\tilde{S}^{(0)}_F(p_\shortparallel;x^d-y^d)-i\eta_R\tilde{S}^{(0)}_F(p_\shortparallel;x^d+y^d-2a_R)\gamma^d&\text{if
		 }~x^d,y^d>a_R
	\end{cases}.
\end{equation}
Using the same reasoning, the propagator in the presence of $N>2$ walls can be constructed from the $N=2$ case.

%%%%%%%%%%%%%%%%%%%%%%%%%%%%%%%%%%%%%%%%%%%%%%%%%%%%%%%%%%%%%%%%%%%%%%%%
%%%%%%%%%%%%%%%%%%%%% New Appendix B %%%%%%%%%%%%%%%%%%%%%%%%%%%%%%%%%%
%%%%%%%%%%%%%%%%%%%%%%%%%%%%%%%%%%%%%%%%%%%%%%%%%%%%%%%%%%%%%%%%%%%%%%%%
\section{Casimir energy for arbitrary couplings}\label{app:energy}

In this Appendix we provide the detailed derivation of the closed-form 
expression for the Casimir energy, valid for arbitrary values of the 
coupling constants $g_L$ and $g_R$, and show how the bag-condition 
limit~$|g_{L,R}|=2$ used in the body of the paper is recovered.

Starting from~(\ref{eq:funct}) and using translation invariance along the $d$ 
parallel directions, the functional determinant reduces to that of an operator 
acting on the $x^d$ coordinate alone. Since 
$V(x^d)=g_L\delta(x^d-a_L)+g_R\delta(x^d-a_R)$ has support on a finite set of 
points, the operator $\mathcal{O}_0^{-1}\mathcal{V}$ is of finite rank, and 
the Fredholm determinant collapses to that of the spinor $\otimes$ wall-index 
matrix already appearing in~(\ref{eq:T}), generalised to arbitrary couplings:
\begin{equation}\label{eq:M-app}
M(p_\shortparallel) \;=\; 
\begin{pmatrix} 
\mathbb{1}+ig_L\,\widetilde{S}_F^{(0)}(p_\shortparallel;0) & 
ig_R\,\widetilde{S}_F^{(0)}(p_\shortparallel;-a) \\[2pt] 
ig_L\,\widetilde{S}_F^{(0)}(p_\shortparallel;+a) & 
\mathbb{1}+ig_R\,\widetilde{S}_F^{(0)}(p_\shortparallel;0)
\end{pmatrix},
\end{equation}
each block being a $2^{\lfloor n/2\rfloor}\times 2^{\lfloor n/2\rfloor}$ 
matrix in spinor space. Modulo the $a$-independent self-energies, the energy 
density is then
\begin{equation}\label{eq:E-detM-app}
\mathcal{E}(a)\;=\;i\int\frac{d^d p_\shortparallel}{(2\pi)^d}\,\log\det 
M(p_\shortparallel).
\end{equation}

\medskip

To make the spinor structure manifest, we use the $\mathrm{sgn}(0)=0$ 
prescription in the free propagator, so that 
$\widetilde{S}_F^{(0)}(p_\shortparallel;0)=\frac{1}{2}\,\hat{\not p}_\shortparallel$, 
and $\widetilde{S}_F^{(0)}(p_\shortparallel;\pm a)=\frac{e^{iaP}}{2}\,
(\hat{\not p}_\shortparallel\mp\gamma^d)$, with the shorthand
\begin{equation}
\hat{\not p}_\shortparallel\;\equiv\;\frac{p_\alpha\gamma^\alpha}
{\sqrt{p_\shortparallel^2}},\qquad 
P\;\equiv\;\sqrt{p_\shortparallel^2}\,.
\end{equation}
The relations
\begin{equation}\label{eq:ph-properties}
\hat{\not p}_\shortparallel^{\,2}\;=\;\mathbb{1},\qquad 
(\gamma^d)^2\;=\;-\mathbb{1},\qquad 
\{\hat{\not p}_\shortparallel,\,\gamma^d\}\;=\;0
\end{equation}
will be used repeatedly. With this notation, (\ref{eq:M-app}) becomes
\begin{equation}\label{eq:M-explicit}
M\;=\; 
\begin{pmatrix} 
\mathbb{1}+\dfrac{ig_L}{2}\,\hat{\not p}_\shortparallel & 
\dfrac{ig_R}{2}\,e^{iaP}(\hat{\not p}_\shortparallel+\gamma^d) \\[10pt] 
\dfrac{ig_L}{2}\,e^{iaP}(\hat{\not p}_\shortparallel-\gamma^d) & 
\mathbb{1}+\dfrac{ig_R}{2}\,\hat{\not p}_\shortparallel
\end{pmatrix}.
\end{equation}

\medskip

To evaluate $\det M$ we use the Schur complement,
\begin{equation}
\det\!\begin{pmatrix} A & B \\ C & D\end{pmatrix}
\;=\;\det A\,\cdot\,\det\!\bigl(D-CA^{-1}B\bigr),
\end{equation}
with $A\equiv\mathbb{1}+(ig_L/2)\hat{\not p}_\shortparallel$. From 
$\hat{\not p}_\shortparallel^{\,2}=\mathbb{1}$, the eigenvalues of 
$\hat{\not p}_\shortparallel$ are $\pm 1$, each with multiplicity 
$2^{\lfloor n/2\rfloor-1}$, hence
\begin{equation}\label{eq:detA-app}
\det A\;=\;\left(1+\frac{g_L^2}{4}\right)^{2^{\lfloor n/2\rfloor-1}}
\;=\;\left[\frac{4+g_L^2}{4}\right]^{2^{\lfloor n/2\rfloor-1}},
\qquad
A^{-1}\;=\;\frac{\mathbb{1}-(ig_L/2)\hat{\not p}_\shortparallel}{1+g_L^2/4}\,.
\end{equation}

The Schur complement involves the product 
$(\hat{\not p}_\shortparallel-\gamma^d)\bigl[\mathbb{1}-(ig_L/2)
\hat{\not p}_\shortparallel\bigr](\hat{\not p}_\shortparallel+\gamma^d)$. 
Using~(\ref{eq:ph-properties}) and the identity 
$\gamma^d\,\hat{\not p}_\shortparallel\,\gamma^d
=-\hat{\not p}_\shortparallel(\gamma^d)^2=\hat{\not p}_\shortparallel$, 
the $g_L$-dependent contribution cancels exactly:
\begin{align}
(\hat{\not p}_\shortparallel-\gamma^d)
(\hat{\not p}_\shortparallel+\gamma^d) &\;=\; 
2\bigl(\mathbb{1}+\hat{\not p}_\shortparallel\gamma^d\bigr)
\;\equiv\;4\,\Pi_+\,, \label{eq:pp-prod}\\[2pt]
(\hat{\not p}_\shortparallel-\gamma^d)\,\hat{\not p}_\shortparallel\,
(\hat{\not p}_\shortparallel+\gamma^d) &\;=\;
\hat{\not p}_\shortparallel-\gamma^d\hat{\not p}_\shortparallel\gamma^d 
\;=\;\hat{\not p}_\shortparallel-\hat{\not p}_\shortparallel\;=\;0\,, 
\label{eq:p-cancel}
\end{align}
where 
\begin{equation}\label{eq:projectors}
\Pi_\pm\;\equiv\;\tfrac{1}{2}\bigl(\mathbb{1}\pm\hat{\not p}_\shortparallel
\gamma^d\bigr)
\end{equation}
are orthogonal projectors associated with the involution 
$\hat{\not p}_\shortparallel\gamma^d$ (which squares to 
$-\hat{\not p}_\shortparallel^{\,2}(\gamma^d)^2=\mathbb{1}$); each has 
rank~$2^{\lfloor n/2\rfloor-1}$. Therefore
\begin{equation}\label{eq:CAinvB}
CA^{-1}B\;=\;\frac{(ig_L/2)(ig_R/2)\,e^{2iaP}}{1+g_L^2/4}\,\cdot\,4\Pi_+
\;=\;-\,\frac{g_L g_R\,e^{2iaP}}{1+g_L^2/4}\,\Pi_+\,,
\end{equation}
so that the Schur complement reads
\begin{equation}\label{eq:Schur}
D-CA^{-1}B\;=\;\mathbb{1}+\frac{ig_R}{2}\,\hat{\not p}_\shortparallel
+\frac{g_L g_R\,e^{2iaP}}{1+g_L^2/4}\,\Pi_+\,.
\end{equation}

\medskip

The operator on the right-hand side of~(\ref{eq:Schur}) belongs to the 
two-dimensional Clifford algebra generated by $\hat{\not p}_\shortparallel$ 
and~$\gamma^d$. This algebra has a single irreducible representation of 
dimension~2; the full spinor space decomposes as 
$2^{\lfloor n/2\rfloor-1}$ copies of it. In any such irreducible block, the 
choice
\begin{equation}
\hat{\not p}_\shortparallel\;\rightarrow\;\sigma_3\,,\qquad 
\gamma^d\;\rightarrow\;i\sigma_2\,,\qquad 
\hat{\not p}_\shortparallel\gamma^d\;\rightarrow\;\sigma_1
\end{equation}
realises~(\ref{eq:ph-properties}) and yields, for the operator 
in~(\ref{eq:Schur}), the matrix
\begin{equation}
\begin{pmatrix} 
1+\dfrac{g_L g_R\,e^{2iaP}}{1+g_L^2/4}\cdot\dfrac{1}{2}(1+1) & 
\dfrac{ig_R}{2} \\[8pt] 
\dfrac{ig_R}{2} & 
1+\dfrac{g_L g_R\,e^{2iaP}}{1+g_L^2/4}\cdot\dfrac{1}{2}(1-1)
\end{pmatrix}
\end{equation}
in the eigenbasis of $\sigma_1$. Its determinant evaluates to
\begin{equation}\label{eq:det-Schur}
\det\!\bigl(D-CA^{-1}B\bigr)\;=\;
\left[\frac{(1+g_L^2/4)(1+g_R^2/4)+g_L g_R\,e^{2iaP}}
{1+g_L^2/4}\right]^{2^{\lfloor n/2\rfloor-1}}.
\end{equation}
Combining~(\ref{eq:detA-app}) and~(\ref{eq:det-Schur}) we arrive at the closed 
form
\begin{equation}\label{eq:detM-final}
\det M(p_\shortparallel)\;=\;
\left[\frac{(4+g_L^2)(4+g_R^2)}{16}\right]^{2^{\lfloor n/2\rfloor-1}}
\!\Bigl[1+\rho_L\rho_R\,e^{2iaP}\Bigr]^{2^{\lfloor n/2\rfloor-1}}\;,
\end{equation}
with the ``reflection coefficients''
\begin{equation}\label{eq:rho-app}
\rho_i\;\equiv\;\frac{4\,g_i}{4+g_i^2}\,,\qquad i=L,R\,.
\end{equation}

\medskip

The first factor in~(\ref{eq:detM-final}) is $a$-independent and accounts for 
the self-energies of the walls; it is removed by the standard subtraction of 
the $a\to\infty$ value of $\mathcal{E}(a)$. Inserting~(\ref{eq:detM-final}) 
in~(\ref{eq:E-detM-app}) and performing the Wick rotation 
($p^0\to ip_E^0$, with the Feynman prescription 
$P=\sqrt{p_\shortparallel^2+i\varepsilon}\to i|p_\shortparallel|$ and 
$i\,d^d p_\shortparallel\to-d^d p_\shortparallel^{E}$), one obtains the 
renormalised energy for arbitrary couplings,
\begin{equation}\label{eq:E-arbitrary}
\mathcal{E}(a)\;=\;-\,2^{\lfloor n/2\rfloor-1}\!\int\!
\frac{d^d p_\shortparallel}{(2\pi)^d}\,
\log\!\bigl[1+\rho_L\rho_R\,e^{-2a|p_\shortparallel|}\bigr],
\end{equation}
which is the formula quoted in the body of the paper. The coefficients 
$\rho_i$ satisfy $|\rho_i|\le 1$, with the upper bound saturated only when 
$|g_i|=2$, where $\rho_i\to\eta_i=\pm1$ and~(\ref{eq:E-arbitrary}) reduces to 
the bag-condition expression used throughout the paper. This realises the 
expectation that the bag-condition value $|g|=2$ is the regime of strongest 
Casimir interaction, in agreement with the explicit results in $1+1$ and 
$3+1$ dimensions of~\cite{FoscoLosada2008}.

%%%%%%%%%%%%%%%%%%%%%%%%%%%%%%%%%%%%%%%%%%%%%%%%%%%%%%%%%%%%%%%%%%%%%%%%%
%%%%%%%%%%%%%%%%%%%%%%%%%%% References %%%%%%%%%%%%%%%%%%%%%%%%%%%%%%%%%%
%%%%%%%%%%%%%%%%%%%%%%%%%%%%%%%%%%%%%%%%%%%%%%%%%%%%%%%%%%%%%%%%%%%%%%%%%

\end{document}